\documentclass[proceedings]{JHEP3}
\usepackage{amsfonts}
\usepackage{amsmath}
\usepackage{epsfig,multicol}

\newbox\mybox

\newcommand\fverb{\setbox\mybox=\hbox\bgroup\verb}
\newcommand\fverbdo{\egroup\medskip\noindent\fbox{\unhbox\mybox}\ }
\newcommand\fverbit{\egroup\item[\fbox{\unhbox\mybox}]}
\conference{Supersymmetric integrable scattering theories with unstable particles}
\abstract{We propose scattering matrices for  N=1 supersymmetric integrable
quantum field theories in 1+1 dimensions which involve unstable particles in their spectra. By 
means of the thermodynamic Bethe ansatz we
analyze the ultraviolet behaviour of some of these theories and identify the
effective Virasoro central charge of the underlying conformal field theories.}

\title{Supersymmetric integrable scattering theories \\
with unstable particles}
\author{Andreas Fring \\
Centre for Mathematical Science, City University, \\
Northampton Square, London EC1V 0HB, UK\\
E-mail: \email{A.Fring@city.ac.uk}}

\input{tcilatex}

\begin{document}

\section{Introduction}

Supersymmetry is a natural concept in particle physics changing bosons into
fermions and vice versa in a well controlled manner. The original idea
traces back over thirty years and its discovery is attributed to Golfand and
Likhtman \cite{Golfand}. Initially the understanding was mainly developed in
the context of string theory \cite{SUSYs1,SUSYs2,SUSYs3}, where
supersymmetry plays the important role of a two-dimensional symmetry of the
world sheet. Thereafter it became a more widely, albeit not universally,
accepted principle when the two dimensional symmetry was generalized to four
dimensions for the Wess-Zumino model \cite{WZ}. Whereas most symmetries lead
to trivial scattering theories, as a consequence of the Coleman-Mandula
theorem \cite{Coleman:1967ad}, supersymmetry is a very special one in the
sense that it can be present and still does not prevent the theory to be
non-trivial. This holds even in higher dimensions. Hence, supersymmetry is
regarded as one of the concepts worthwhile studying in 1+1 dimensions as it
might even have a direct bearing on higher dimensional theories.

The first scattering matrices for N=1 supersymmetric integrable quantum
field theories in 1+1 dimensions were constructed by Witten and Shankar in
the late seventies \cite{Shankar} for theories with degenerate mass spectra.
In a systematic manner these results were generalized by Schoutens twelve
years later to theories with non-degenerate mass spectra \cite{Schoutens},
see also \cite{Fendley,Ahn,Evans,Hollowood,Gandenberger,Mussardo}. $S$%
-matrices for some specific cases of $N=2$ supersymmetric theories were
constructed by K{\"{o}}berle and Kurak \cite{N2SUSY} in the late eighties
and thereafter generalized \cite{Fendley2,Kobayashi,Fendley3}. Up to now all
the constructed $S$-matrices invariant under supersymmetry involve
exclusively stable particles in their spectra.

Even though it has been commented upon the occurrence of unstable particles
in integrable quantum field theories in 1+1 dimensions as early as the late
seventies \cite{Z4}, only recently such type of theories have been
investigated in more detail. For instance, scattering matrices for such type
of theories have been constructed \cite{S16,S5,S15}, their ultraviolet
behaviour has been analyzed by means of the thermodynamic Bethe ansatz \cite%
{HSGTBA,D1,S1,vac,D2} and also form factors have been constructed which were
used to compute correlation functions needed in various quantities \cite%
{F8,F6,F7,F5}.

The purpose of this note is to propose and analyze a class of scattering
matrices which are invariant under supersymmetry transformations and contain
besides stable particles also unstable particles in their spectra.

The manuscript is organized as follows: In the next section we briefly
recall some of the main features of $N=1$ supersymmetry relevant for the
development of a scattering theory. In section 3 we present the bootstrap
construction for supersymmetric theories and in section 4 a proposal which
implements in addition unstable particles into such type of theories. In
section 5 we investigate the ultraviolet limit for some of the proposed $S$%
-matrices. The conclusions are stated in section 6.

\section{N=1 supersymmetry, generalities}

We commence by fixing the notation and briefly recall the key features of \ $%
N=1$ supersymmetry which will be relevant below. The prerequisite for a
supersymmetric theory is the existence of two conserved supercharges $%
\mathcal{Q}$ and $\mathcal{\bar{Q}}$ together with a fermion parity operator 
$\mathcal{\hat{Q}}$. The $N=1$ superalgebra\textit{\ }obeyed by these
charges reads in general 
\begin{equation}
\{\mathcal{\hat{Q}},\mathcal{Q}\}=\{\mathcal{\hat{Q}},\mathcal{\bar{Q}}%
\}=0,\quad \mathcal{Q}^{2}=\mathcal{P}_{1},\quad \mathcal{\bar{Q}}^{2}=%
\mathcal{P}_{-1},\quad \mathcal{\hat{Q}}^{2}=\mathbb{I\quad }\text{and\quad }%
\{\mathcal{Q},\mathcal{\bar{Q}}\}=T,  \label{N1}
\end{equation}%
where $\mathcal{P}_{\pm 1}$ are charges of Lorentz spin $\pm 1$ and $T$ is
the topological charge operator. Here we restrict ourselves to the case $T=0$
(see \cite{Schoutens} for some examples with $T\neq 0$). Next we note that
in a massive $N=1$ supersymmetric theory one can arrange all particles in
multiplets $(b_{i},f_{i})$ with internal quantum numbers $1\leq i\leq \ell $
containing a boson $b_{i}$ and a fermion $f_{i}$ with equal masses $%
m_{b_{i}}=$ $m_{f_{i}}=$ $m_{_{i}}$. An asymptotic boson or fermion is
characterized by a creation operator $Z_{\mu _{j}}(\theta )$ depending on
the rapidity $\theta $, which parameterizes the momenta as $p^{0}=m\cosh
\theta $, $p^{1}=m\sinh \theta $. In agreement with (\ref{N1}), one \cite%
{Shankar,Schoutens} can specify the action of the charges on asymptotic
n-particle states. In particular, on a one-particle asymptotic state they
act as 
\begin{eqnarray}
\mathcal{Q}\left\vert Z_{\mu _{j}}(\theta )\right\rangle _{\text{in/out}} &=&%
\sqrt{m_{j}}e^{\theta /2}\left\vert Z_{\hat{\mu}_{j}}(\theta )\right\rangle
_{\text{in/out}}~\qquad \qquad ~~~~~~\text{for \ }\mu =b,f;~~1\leq i\leq
\ell , \\
\mathcal{\bar{Q}}\left\vert Z_{\mu _{j}}(\theta )\right\rangle _{\text{in/out%
}} &=&i(-1)^{F_{\mu _{j}}}\sqrt{m_{j}}e^{\theta /2}\left\vert Z_{\hat{\mu}%
_{j}}(\theta )\right\rangle _{\text{in/out}}~\qquad \text{for \ }\mu
=b,f;~~1\leq i\leq \ell ,~~~~~ \\
\mathcal{\hat{Q}}\left\vert Z_{\mu _{j}}(\theta )\right\rangle _{\text{in/out%
}} &=&(-1)^{F_{\mu _{j}}}\left\vert Z_{\mu _{j}}(\theta )\right\rangle _{%
\text{in/out}}\qquad \qquad \qquad ~~\text{for \ }\mu =b,f;~~1\leq i\leq
\ell ,
\end{eqnarray}%
where we defined 
\begin{equation}
\hat{\mu}=\QATOPD\{ . {b\quad ~\text{for }\mu =f}{f\quad \text{for }\mu
=b}\quad \quad \text{and\quad \quad }F_{\mu }=\QATOPD\{ . {1\quad \text{for }%
\mu =f}{0\quad \text{for }\mu =b}~.
\end{equation}%
For the generalization to an action on n-particle asymptotic states one
defines 
\begin{equation}
\mathcal{Q}^{(n)}=\sum\limits_{k=1}^{n}\left( \bigotimes\limits_{l=1}^{k-1}%
\mathcal{\hat{Q}}_{l}\right) \mathcal{Q}_{k},\quad \mathcal{\bar{Q}}%
^{(n)}=\sum\limits_{k=1}^{n}\left( \bigotimes\limits_{l=1}^{k-1}\mathcal{%
\hat{Q}}_{l}\right) \mathcal{\bar{Q}}_{k},\quad \mathcal{\hat{Q}}%
^{(n)}=\bigotimes\limits_{k=1}^{n}\mathcal{\hat{Q}}_{k}~.
\end{equation}%
As we want to construct two-particle scattering amplitudes, we require in
particular the action of these charges on two-particle states. From the
above definitions one obtains 
\begin{eqnarray}
\mathcal{\hat{Q}}^{(2)}\left\vert Z_{\mu _{j}}(\theta )Z_{\nu _{k}}(\theta
^{\prime })\right\rangle _{\text{in/out}} &=&e^{i\pi (F_{\mu _{j}}+F_{\nu
_{k}})}\left\vert Z_{\mu _{j}}(\theta )Z_{\nu _{k}}(\theta ^{\prime
})\right\rangle _{\text{in/out}}  \notag  \label{Q21} \\
\mathcal{Q}^{(2)}\left\vert Z_{\mu _{j}}(\theta )Z_{\nu _{k}}(\theta
^{\prime })\right\rangle _{\text{in/out}} &=&\sqrt{m_{j}}e^{\frac{\theta }{2}%
}\left\vert Z_{\hat{\mu}_{j}}(\theta )Z_{\nu _{k}}(\theta ^{\prime
})\right\rangle _{\text{in/out}}+\sqrt{m_{k}}e^{\frac{\theta ^{\prime }}{2}%
}\left\vert Z_{\mu _{j}}(\theta )Z_{\hat{\nu}_{k}}(\theta ^{\prime
})\right\rangle _{\text{in/out}}  \notag \\
\mathcal{\bar{Q}}^{(2)}\left\vert Z_{\mu _{j}}(\theta )Z_{\nu _{k}}(\theta
^{\prime })\right\rangle _{\text{in/out}} &=&i\sqrt{m_{j}}e^{\frac{\theta }{2%
}+i\pi F_{\mu _{j}}}\left\vert Z_{\hat{\mu}_{j}}(\theta )Z_{\nu _{k}}(\theta
^{\prime })\right\rangle _{\text{in/out}}  \notag \\
&&+i\sqrt{m_{k}}e^{\frac{\theta ^{\prime }}{2}+i\pi (F_{\mu _{j}}+F_{\nu
_{k}})}\left\vert Z_{\mu _{j}}(\theta )Z_{\hat{\nu}_{k}}(\theta ^{\prime
})\right\rangle _{\text{in/out}}.\ \ \ ~~\ \ ~  \label{Q23}
\end{eqnarray}%
This is sufficient information to determine the consequences on the
scattering theory of an integrable quantum field theory when demanding it to
be supersymmetric.

\section{The bootstrap construction}

Next recall briefly the key steps of the bootstrap construction carried out
by Schoutens \cite{Schoutens} with some minor differences. As a general
structure for $S$ one assumes usually \cite{Shankar,Schoutens} the
factorization into a purely bosonic part $\hat{S}$ and a factor $\check{S}$
which incorporates the boson-fermion mixing. Hence 
\begin{equation}
\left\vert Z_{\alpha _{i}}(\theta )Z_{\beta _{j}}(\theta ^{\prime
})\right\rangle _{\text{in}}=\sum\limits_{k,l=1}^{\ell }\sum\limits_{\gamma
,\delta =b,f}\hat{S}_{ij}^{kl}(\theta -\theta ^{\prime })\check{S}_{\alpha
_{i}\beta _{j}}^{\gamma _{k}\delta _{l}}(\theta -\theta ^{\prime
})\left\vert Z_{\gamma _{k}}(\theta ^{\prime })Z_{\delta _{l}}(\theta
)\right\rangle _{\text{out}}.  \label{Sfact}
\end{equation}
We use here the notation that the Latin indices are the internal quantum
numbers relating to the particle type and the Greek indices distinguish
fermions $f$ from bosons $b$. This means by construction $\hat{S}$ commutes
trivially with all supercharges $\mathcal{Q}$, $\mathcal{\bar{Q}}$ and $%
\mathcal{\hat{Q}}$, whereas the requirement of invariance under
supersymmetry constrains only the boson-fermion mixing factor $\check{S}$.
It seems there is no compelling reason for demanding the factorization (\ref%
{Sfact}) and one could envisage more general constructions, but we follow
here \cite{Shankar,Schoutens} and take $S=\hat{S}\check{S}$ as a working
hypothesis. One can now invoke consecutively the consistency equations from
the bootstrap program \cite{Schroer3,KTTW,ZamS,K2} in order to determine the
precise form of $S$.

\subsection{Constraints from supersymmetry}

Following the most systematic treatment \cite{Schoutens}, one can first of
all analyze the constraints resulting from the requirement that the theory
should be supersymmetric, which means the supercharges should commute with
the scattering matrix. Paying attention to the fact that $S$ intertwines the
asymptotic states it follows with (\ref{Q23}), that the boson-fermion mixing
matrix $\check{S}$ has to obey 
\begin{eqnarray}
\left[ \hat{Q}\otimes \hat{Q}\right] \check{S}(\theta _{12}) &=&~\check{S}%
(\theta _{12})\left[ \hat{Q}\otimes \hat{Q}\right]  \label{c1} \\
\left[ m_{k}^{\frac{1}{2}}e^{\frac{\theta _{2}}{2}}Q\otimes \mathbb{I+}%
m_{j}^{\frac{1}{2}}e^{\frac{\theta _{1}}{2}}\hat{Q}\otimes Q\right] \check{S}%
(\theta _{12}) &=&\check{S}(\theta _{12})\left[ m_{j}^{\frac{1}{2}}e^{\frac{%
\theta _{1}}{2}}Q\otimes \mathbb{I+}m_{k}^{\frac{1}{2}}e^{\frac{\theta _{2}}{%
2}}\hat{Q}\otimes Q\right]  \label{c2} \\
\left[ m_{k}^{\frac{1}{2}}e^{\frac{-\theta _{2}}{2}}\bar{Q}\otimes \mathbb{I+%
}m_{j}^{\frac{1}{2}}e^{\frac{-\theta _{1}}{2}}\hat{Q}\otimes \bar{Q}\right] 
\check{S}(\theta _{12}) &=&\check{S}(\theta _{12})\left[ m_{j}^{\frac{1}{2}%
}e^{\frac{-\theta _{1}}{2}}\bar{Q}\otimes \mathbb{I+}m_{k}^{\frac{1}{2}}e^{%
\frac{-\theta _{2}}{2}}\hat{Q}\otimes \bar{Q}\right] .~~~~  \label{c3}
\end{eqnarray}
As usual we abbreviated the rapidity difference $\theta _{12}:=\theta
_{1}-\theta _{2}$. With the explicit realization \cite{Shankar,Schoutens}
for the $N=1$-superalgebra 
\begin{equation}
Q=\left( 
\begin{array}{ll}
0 & 1 \\ 
1 & 0%
\end{array}
\right) ,\qquad \bar{Q}=\left( 
\begin{array}{rr}
0 & -i \\ 
i & 0%
\end{array}
\right) \quad \text{and\quad }\hat{Q}=\left( 
\begin{array}{rr}
1 & 0 \\ 
0 & -1%
\end{array}
\right) ,
\end{equation}
one can solve (\ref{c1})-(\ref{c3}). First one notices that (\ref{c1})
implies that the fermion parity has to be the same in the in- and out-state
such that only the eight processes $bb\rightarrow (bb,ff)$, $bf\rightarrow
(bf,fb)$, $fb\rightarrow (bf,fb)$ and $ff\rightarrow (bb,ff)$ can possibly
be non-vanishing. Invoking then also the relations (\ref{c2})-(\ref{c3})
fixes the $S$-matrix up to two entries. Instead of leaving two amplitudes
unknown at this stage, it is convenient to introduce \cite{Schoutens} two
unknown functions $f_{ij}(\theta )$ and $g_{ij}(\theta )$. These functions
carry only two indices when one makes the further assumption that the
bosonic $S$-matrix describes a theory with a non-degenerate mass spectrum
such that backscattering is absent and $\hat{S}$ is diagonal in the sense of 
$\hat{S}_{ij}^{kl}(\theta )=\delta _{i}^{l}\delta _{j}^{k}\hat{S}%
_{ij}(\theta )$. Even though, when taking the bosonic factor of $S$ to be
diagonal, the mass degeneracy between bosons and fermions of the same type
forces $\check{S}$ to be of the form 
\begin{equation}
\check{S}_{ij}(\theta )=\left( 
\begin{array}{llll}
\check{S}_{b_{i}b_{j}}^{b_{j}b_{i}}(\theta ) & 0 & 0 & \check{S}%
_{b_{i}b_{j}}^{f_{j}f_{i}}(\theta ) \\ 
0 & \check{S}_{b_{i}f_{j}}^{b_{j}f_{i}}(\theta ) & \check{S}%
_{b_{i}f_{j}}^{f_{j}b_{i}}(\theta ) & 0 \\ 
0 & \check{S}_{f_{i}b_{j}}^{b_{j}f_{i}}(\theta ) & \check{S}%
_{f_{i}b_{j}}^{f_{j}b_{i}}(\theta ) & 0 \\ 
\check{S}_{f_{i}f_{j}}^{b_{j}b_{i}}(\theta ) & 0 & 0 & \check{S}%
_{f_{i}f_{j}}^{f_{j}f_{i}}(\theta )%
\end{array}
\right) ,~\ \ \ \ \ \ \ \ \text{for \ }1\leq i,j\leq \ell .~~~~
\end{equation}
To avoid the occurrence of additional phase factors one can include them
directly into the asymptotic states and change $Z_{f_{j}}(\theta
)\rightarrow \exp (-i\pi /4)Z_{f_{j}}(\theta )$. In this new basis Schoutens
found \cite{Schoutens} as solutions to (\ref{c1})-(\ref{c3}) 
\begin{equation}
\check{S}_{ij}(\theta )=\frac{2f_{ij}(\theta )}{\rho _{ij}^{+}+\cosh \frac{%
\theta }{2}}\left( 
\begin{array}{rrrr}
\rho _{ij}^{+} & 0 & 0 & -i\sinh \frac{\theta }{2} \\ 
0 & \cosh \frac{\theta }{2} & -\rho _{ij}^{-} & 0 \\ 
0 & \rho _{ij}^{-} & \cosh \frac{\theta }{2} & 0 \\ 
-i\sinh \frac{\theta }{2} & 0 & 0 & \rho _{ij}^{+}%
\end{array}
\right) +g_{ij}(\theta )\left( 
\begin{array}{rrrr}
1 & 0 & 0 & 0 \\ 
0 & 0 & 1 & 0 \\ 
0 & 1 & 0 & 0 \\ 
0 & 0 & 0 & -1%
\end{array}
\right) ~~~~~  \label{S1234}
\end{equation}
with $\rho _{ij}^{\pm }=\left[ (m_{i}/m_{j})^{1/2}\pm (m_{j}/m_{i})^{1/2}%
\right] /2$. One observes that the requirement of supersymmetry invariance
puts severe constraints on the general structure of the $S$-matrix, albeit
it does not fix it entirely. Thus leaving fortunately enough freedom to
incorporate also other necessary features.

\subsection{Constraints from the Yang-Baxter equations}

Next we invoke the equations which are the consequence of the
factorizability of the $n$-particle $S$-matrix into two particle scattering
amplitudes. Since we have mass degeneracy between bosons and fermions of the
same type backscattering is possible and the Yang-Baxter equations \cite%
{Yang,Baxter} 
\begin{equation}
\sum\limits_{\kappa _{1},\kappa _{2},\kappa _{3}}S_{\mu _{1}\mu
_{2}}^{\kappa _{1}\kappa _{2}}(\theta _{12})S_{\kappa _{2}\mu _{3}}^{\kappa
_{3}\nu _{1}}(\theta _{13})S_{\kappa _{1}\kappa _{3}}^{\nu _{3}\nu
_{2}}(\theta _{23})=\sum\limits_{\kappa _{1},\kappa _{2},\kappa _{3}}S_{\mu
_{2}\mu _{3}}^{\kappa _{1}\kappa _{2}}(\theta _{23})S_{\mu _{1}\kappa
_{1}}^{\nu _{3}\kappa _{3}}(\theta _{13})S_{\kappa _{3}\kappa _{2}}^{\nu
_{2}\nu _{1}}(\theta _{12})  \label{YB}
\end{equation}%
will impose further non-trivial constraints on $S$. It was noted in \cite%
{Schoutens}, that in order to satisfy (\ref{YB}) with (\ref{S1234}) one can
fix the ratio between the functions  $f_{ij}(\theta )$ and $g_{ij}(\theta )$
up to an unknown constant $\kappa $%
\begin{equation}
f_{ij}(\theta )=\frac{\kappa \sqrt{m_{i}m_{j}}}{2}\left( \frac{\rho
_{ij}^{+}+\cosh \frac{\theta }{2}}{\sinh \theta }\right) g_{ij}(\theta )~,
\label{fg}
\end{equation}%
such that 
\begin{equation}
\!\check{S}_{ij}(\theta )=g_{ij}(\theta )\left[ \frac{\kappa \sqrt{m_{i}m_{j}%
}}{\sinh \theta }\left( 
\begin{array}{rrrr}
\rho _{ij}^{+} & 0 & 0 & -i\sinh \frac{\theta }{2} \\ 
0 & \cosh \frac{\theta }{2} & -\rho _{ij}^{-} & 0 \\ 
0 & \rho _{ij}^{-} & \cosh \frac{\theta }{2} & 0 \\ 
-i\sinh \frac{\theta }{2} & 0 & 0 & \rho _{ij}^{+}%
\end{array}%
\right) +\left( 
\begin{array}{rrrr}
1 & 0 & 0 & 0 \\ 
0 & 0 & 1 & 0 \\ 
0 & 1 & 0 & 0 \\ 
0 & 0 & 0 & -1%
\end{array}%
\right) \right] \!.  \label{S2}
\end{equation}%
If we were dealing with a lattice model we would have already solved the
problem to find a consistent supersymmetric $R$-matrix in that case.
However, aiming at the description of a quantum field theory we also have to
incorporate all the analytic properties.

\subsection{Constraints from hermitian analyticity, unitarity and crossing}

\noindent A scattering matrix belonging to a proper quantum field theory has
to be hermitian analyticity \cite{David,LM}, unitarity and crossing
invariant \cite{book,Schroer3,KTTW,ZamS} 
\begin{equation}
S_{ij}^{kl}(\theta _{ij})=\left[ S_{kl}^{ij}(-\theta _{ij}^{\ast })\right]
^{\ast },~~~\sum_{kl}S_{ij}^{kl}(\theta )\left[ S_{nm}^{kl}(\theta )\right]
^{\ast }=\delta _{in}\delta _{jm},~~S_{ij}^{kl}(\theta _{ij})=S_{\bar{k}i}^{l%
\bar{\jmath}}(i\pi -\theta _{ij}).
\end{equation}
It is easy to convince oneself that hermitian analyticity and crossing are
satisfied when $\kappa \in \mathbb{R}$ and in addition 
\begin{equation}
g_{kj}(\theta )=g_{jk}^{\ast }(-\theta )\qquad \text{and\qquad }%
g_{kj}(\theta )=g_{\bar{k}j}(i\pi -\theta )  \label{con}
\end{equation}
hold. The unitarity requirement is satisfied once we fulfill the functional
relation 
\begin{equation}
g_{ij}(\theta )g_{ji}(-\theta )=\left[ 1-\kappa ^{2}m_{i}m_{j}\left( \frac{%
\left( \rho _{ij}^{+}\right) ^{2}+\sinh ^{2}\frac{\theta }{2}}{\sinh
^{2}\theta }\right) \right] ^{-1}~=:\chi _{ij}(\theta ).  \label{un}
\end{equation}
In order to solve the set of equations (\ref{con})-(\ref{un}) we assume now
first parity invariance for $g$ and self-conjugacy for the particles
involved 
\begin{equation}
g_{kj}(\theta )=g_{jk}(\theta )\qquad \text{and\qquad }g_{jk}(\theta )=g_{j%
\bar{k}}(\theta ).  \label{zu}
\end{equation}
Following a standard procedure to solve functional equations of the above
type we make the general ansatz 
\begin{equation}
g_{ij}(\theta )=\lambda \prod\limits_{l=1}^{\infty }\frac{\rho _{ij}[\theta
+2\pi il]\rho _{ij}[-\theta +2\pi i(l+1/2)]}{\rho _{ij}[\theta +2\pi
i(l+1/2)]\rho _{ij}[-\theta +2\pi i(l+1)]}.  \label{ans}
\end{equation}
At this stage $\lambda \in \mathbb{C}$ is some arbitrary constant and the $%
\rho _{ij}$ are some functions which still need to be determined. The ansatz
(\ref{ans}) solves the crossing relation (\ref{con}) by construction when
also (\ref{zu}) holds. Substituting (\ref{ans}) into (\ref{un}) we then find
that 
\begin{equation}
\chi _{ij}(\theta )=\lambda ^{2}\rho _{ij}\left( \theta +2\pi i\right) \rho
_{ij}\left( 2\pi i-\theta \right)  \label{rho}
\end{equation}
has to be satisfied. Hence, we have reduced the problem of simultaneously
solving (\ref{con}) and (\ref{un}) to a much simpler problem of just
factorizing the function $\chi $. Unfortunately, (\ref{un}) can not yet be
compared directly with (\ref{rho}), but it was noted in \cite{Schoutens}
that when introducing two auxiliary equations which parameterize the masses
and $\kappa $ in terms of the new quantities $\eta _{ij}$, $\hat{\eta}_{ij}$ 
\begin{equation}
\frac{\kappa ^{2}}{2}m_{i}m_{j}=\cos \eta _{ij}+\cos \hat{\eta}_{ij}\quad 
\text{and\quad }-\frac{\kappa ^{2}}{4}\left( m_{i}^{2}+m_{j}^{2}\right)
=1+\cos \eta _{ij}\cos \hat{\eta}_{ij},  \label{para}
\end{equation}
one can bring $\chi $ into a more suitable form 
\begin{equation}
\chi _{ij}(\theta )=\frac{\sinh ^{2}\frac{\theta }{2}\cosh ^{2}\frac{\theta 
}{2}}{\sinh \frac{1}{2}\left( \theta +i\eta _{ij}\right) \sinh \frac{1}{2}%
\left( \theta -i\eta _{ij}\right) \sinh \frac{1}{2}\left( \theta +i\hat{\eta}%
_{ij}\right) \sinh \frac{1}{2}\left( \theta -i\hat{\eta}_{ij}\right) }.
\label{xfact}
\end{equation}
\noindent Comparing now (\ref{rho}) and (\ref{xfact}) there are obviously
various solutions. Starting by producing the factors $\pi ^{2}/\sinh \frac{1%
}{2}\left( \theta +i\eta \right) \sinh \frac{1}{2}\left( \theta -i\eta
\right) $ for $\eta =\eta _{ij}$,$\hat{\eta}_{ij}$, we have the
possibilities 
\begin{eqnarray}
\rho _{ij}^{(1/2)}\left( \theta +2\pi i,\eta \right) &=&\Gamma \left( \frac{%
i\theta \mp \eta }{2\pi }\right) \Gamma \left( 1+\frac{i\theta \pm \eta }{%
2\pi }\right) ,\quad  \label{r1} \\
\rho _{ij}^{(3/4)}\left( \theta +2\pi i,\eta \right) &=&\Gamma \left( \frac{%
-i\theta \mp \eta }{2\pi }\right) \Gamma \left( 1-\frac{i\theta \mp \eta }{%
2\pi }\right) , \\
\rho _{ij}^{(5/6)}\left( \theta +2\pi i,\eta \right) &=&\pm \pi /\sinh \frac{%
1}{2}\left( \theta \pm i\eta \right) .  \label{r3}
\end{eqnarray}
We can now substitute these solutions back into (\ref{ans}) in order to
assemble $g_{ij}(\theta ,\eta )$. When restricting w.l.g. the parameters $%
0<\eta _{ij},\hat{\eta}_{ij}<\pi $, we observe that all functions $%
g_{ij}(\theta ,\eta )$ have poles inside the physical sheet, that is $0<%
\limfunc{Im}\theta \leq \pi $, except the one constructed from $\rho
_{ij}^{(4)}\left( \theta ,\eta \right) $. Thus only for this solution the
boson-fermion mixing factor $\check{S}$ does not introduce new bound states
(see next subsection for more details on fusing), such that the fusing
structure is entire contained in the bosonic factor $\hat{S}$. Selecting out
this particular solution we can write 
\begin{equation}
g_{ij}(\theta )=\frac{1}{2i}\frac{\sinh \theta }{\sinh \frac{1}{2}(\theta
+i\eta _{ij})\sinh \frac{1}{2}(\theta +i\hat{\eta}_{ij})}g(\theta ,\eta
_{ij})g(\theta ,\hat{\eta}_{ij})
\end{equation}
where we defined the function 
\begin{eqnarray}
g(\theta ,\eta ) &=&i\prod\limits_{k=1}^{\infty }\frac{\Gamma \left( k-\frac{%
i\theta +\eta }{2\pi }\right) \Gamma \left( k+\frac{1}{2}+\frac{i\theta
-\eta }{2\pi }\right) \Gamma \left( k-\frac{i\theta -\eta }{2\pi }\right)
\Gamma \left( k-\frac{1}{2}+\frac{i\theta +\eta }{2\pi }\right) }{\Gamma
\left( k+\frac{i\theta -\eta }{2\pi }\right) \Gamma \left( k+\frac{1}{2}-%
\frac{i\theta +\eta }{2\pi }\right) \Gamma \left( k+\frac{i\theta +\eta }{%
2\pi }\right) \Gamma \left( k-\frac{1}{2}-\frac{i\theta -\eta }{2\pi }%
\right) }  \notag \\
&=&\exp \left[ \int\nolimits_{0}^{\infty }\frac{dt}{t}\left[ \frac{\sinh
t\left( \frac{1}{2}-\frac{\eta }{\pi }\right) }{2\sinh \frac{t}{2}\cosh
^{2}t/2}-1\right] \sinh \frac{t\theta }{2\pi i}\right] .  \label{gint}
\end{eqnarray}
Clearly it would be very interesting to investigate also the solutions
resulting from the functions (\ref{r1})-(\ref{r3}) other than $\rho
_{ij}^{(4)}$. Further solutions can be expected when one relaxes the
assumptions (\ref{zu}).

\subsection{Constraints from the boundstate bootstrap equations}

The last remaining constraint arises when we consider the consequences of
the factorization of the $S$-matrix in conjunction with the possibility of a
fusing process, say $\mu _{i}+\nu _{j}\rightarrow $ $\bar{\kappa}_{k}$, for $%
\mu ,\nu ,\kappa =b,f$ and $1\leq i,j,k\leq \ell $. For this to happen the
scattering matrix must posses a simple order pole in the physical sheet at
some fusing angle $i\eta _{\mu _{i}\nu _{j}}^{\bar{\kappa}_{k}}$ with $\eta
_{\mu _{i}\nu _{j}}^{\bar{\kappa}_{k}}\in \mathbb{R}^{+}$. The residue of $S$
at this angle is related to the three-point couplings $\Gamma _{\mu _{i}\nu
_{j}}^{\bar{\kappa}_{k}}$ via 
\begin{equation}
i\limfunc{Res}_{\theta \rightarrow i\eta _{\mu _{i}\nu _{j}}^{\bar{\kappa}%
_{k}}}S_{\mu _{i}\nu _{j}}^{\rho _{l}\tau _{m}}(\theta )=\sum\limits_{\bar{%
\kappa}_{k}}\left( \Gamma _{\rho _{l}\tau _{m}}^{\bar{\kappa}_{k}}\right)
^{\ast }\Gamma _{\mu _{i}\nu _{j}}^{\bar{\kappa}_{k}}  \label{Res}
\end{equation}
Then the following boundstate bootstrap equation \cite%
{Schroer3,SUN2,Karobound,K2} 
\begin{equation}
\sum\limits_{\delta _{d},\gamma _{g},\rho _{n}}\Gamma _{\delta _{d}\rho
_{n}}^{\beta _{b}}S_{\mu _{i}\gamma _{g}}^{\alpha _{a}\delta _{d}}(\theta +i%
\bar{\eta}_{\kappa _{k}\mu _{i}}^{\bar{\nu}_{j}})S_{\nu _{j}\lambda
_{l}}^{\gamma _{g}\rho _{n}}(\theta -i\bar{\eta}_{\nu _{j}\kappa _{k}}^{\bar{%
\mu}_{i}})=\sum\limits_{\bar{\kappa}_{k}}S_{\bar{\kappa}_{k}\lambda
_{l}}^{\alpha _{a}\beta _{b}}(\theta )\Gamma _{\mu _{i}\nu _{j}}^{\bar{\kappa%
}_{k}}  \label{bbe}
\end{equation}
has to be satisfied. Here the $\bar{\eta}$ is related to the fusing angle $%
\eta $ as $\bar{\eta}=\pi -\eta $. Taking the factorization ansatz (\ref%
{Sfact}) for $S$ into account and assuming further that the bosonic part of
the scattering matrix is diagonal $\hat{S}_{ij}^{kl}(\theta )=\delta
_{i}^{l}\delta _{j}^{k}\hat{S}_{ij}(\theta )$, the relation (\ref{bbe})
simplifies to 
\begin{equation}
\sum\limits_{\delta ,\gamma ,\rho }\Gamma _{\delta _{i}\rho _{j}}^{\beta
_{b}}S_{\mu _{i}\gamma _{l}}^{\alpha _{l}\delta _{i}}(\theta +i\bar{\eta}%
_{\kappa _{k}\mu _{i}}^{\bar{\nu}_{j}})S_{\nu _{j}\lambda _{l}}^{\gamma
_{l}\rho _{j}}(\theta -i\bar{\eta}_{\nu _{j}\kappa _{k}}^{\bar{\mu}%
_{i}})=\sum\limits_{\bar{\kappa}}S_{\bar{\kappa}_{b}\lambda _{l}}^{\alpha
_{l}\beta _{b}}(\theta )\Gamma _{\mu _{i}\nu _{j}}^{\bar{\kappa}_{b}}
\label{bbb}
\end{equation}
It is not difficult to convince oneself that (\ref{bbb}) results from the
formal equation 
\begin{equation}
Z_{\mu _{i}}\left( \theta +i\bar{\eta}_{\mu _{k}\mu _{i}}^{\bar{\mu}%
_{j}}\right) Z_{\nu _{j}}\left( \theta -i\bar{\eta}_{\mu _{j}\mu _{k}}^{\bar{%
\mu}_{i}}\right) =\sum\limits_{\bar{\kappa}_{k}}\Gamma _{\mu _{i}\nu _{j}}^{%
\bar{\kappa}_{k}}Z_{\bar{\kappa}_{k}}(\theta ).  \label{fuse}
\end{equation}
together with the assumption that the $Z$s obey a Zamolodchikov algebra \cite%
{ZZ}, i.e. when exchanging (braiding) them they will pick up an $S$-matrix
as a structure constant. Acting on (\ref{fuse}) with $\mathcal{\hat{Q}}$ one
notices first of all that only the following fusing processes are allowed to
occur 
\begin{equation}
b_{i}+b_{j}\rightarrow b_{k},\quad f_{i}+f_{j}\rightarrow b_{k},\quad
b_{i}+f_{j}\rightarrow f_{k},\quad f_{i}+b_{j}\rightarrow f_{k}.
\end{equation}
Furthermore when acting with $\mathcal{Q}$ and $\mathcal{\bar{Q}}$ on (\ref%
{fuse}) one finds a powerful constraint for the three point couplings 
\begin{equation}
\left( \frac{\Gamma _{b_{i}b_{j}}^{b_{k}}}{\Gamma _{f_{i}f_{j}}^{b_{k}}}%
\right) ^{2}=\frac{m_{k}+m_{i}+m_{j}}{m_{i}+m_{j}-m_{k}}.
\end{equation}
Computing then the residues of $S$ by means of (\ref{Res}) for the processes 
$b_{i}+b_{j}\rightarrow b_{k}$ and $f_{i}+f_{j}\rightarrow b_{k}$ yields for
the constant $\kappa $ in (\ref{S2}) the relation 
\begin{equation}
\kappa =\frac{\sin \eta _{ij}^{k}}{\sqrt{m_{i}m_{j}}\rho _{ij}^{+}}\left[ 
\frac{\left( \Gamma _{b_{i}b_{j}}^{b_{k}}\right) ^{2}+\left( \Gamma
_{f_{i}f_{j}}^{b_{k}}\right) ^{2}}{\left( \Gamma
_{b_{i}b_{j}}^{b_{k}}\right) ^{2}-\left( \Gamma _{f_{i}f_{j}}^{b_{k}}\right)
^{2}}\right] =\frac{2\sin \eta _{ij}^{k}}{m_{k}}~.  \label{kconst}
\end{equation}
Notice that this is quite a severe constraint as the right hand side of (\ref%
{kconst}) has to hold universally for all possible values of $i,j,k$.

\section{Implementing unstable particles}

As a consequence of the factorizing ansatz (\ref{Sfact}) for $S$ and the
choice for $\check{S}$ which does not possess poles inside the physical
sheet, the pole structure responsible for fusing processes is entirely
confined to the bosonic factor $\hat{S}$. One may therefore search the large
reservoir of diagonal $S$-matrices to find suitable solutions. In the
original paper Schoutens \cite{Schoutens} noticed that one may satisfy (\ref%
{kconst}) with $\hat{S}$ equal to the scattering matrix of minimal $A_{2\ell
}^{(2)}$-affine Toda field theory \cite{FKM} 
\begin{equation}
S_{ab}(\theta )=\left( \frac{a+b}{2n+1}\right) _{\theta }\,\left( \frac{|a-b|%
}{2n+1}\right) _{\theta }\prod_{k=1}^{\min (a,b)-1}\left( \frac{a+b-2k}{2n+1}%
\right) _{\theta }^{2}\,\,  \label{sfact}
\end{equation}
for $1\leq a,b\leq \ell $ and with $(x)_{\theta }:=\tanh \frac{1}{2}(\theta
+i\pi x)/\tanh \frac{1}{2}(\theta -i\pi x)$. Thereafter, Hollowood and
Mavrikis \cite{Hollowood} showed that when taking the bosonic factor $\hat{S}
$ to be the minimal $A_{\ell }^{(1)}$, $D_{\ell }^{(1)}$ or $(C_{\ell
}^{(1)}|D_{\ell +2}^{(2)})$-affine Toda $S$-matrix, the ansatz (\ref{Sfact})
also satisfies (\ref{kconst}) together with the bootstrap equations, thus
leading to consistent supersymmetric $S$-matrices.

Based on these results it is straightforward to extend the ansatz and also
include unstable particles into the spectrum of these theories. We may take
the bosonic factor of $\hat{S}$ to belong to the large class of models which
can be referred to conveniently as $\mathbf{g|\tilde{g}}$-theories. In these
models each particle carries two quantum numbers $(a,i)$, one associated to
the algebra $\mathbf{g}$ with $1\leq a\leq \ell =$ rank$\mathbf{g}$ and the
other related to the algebra $\mathbf{\tilde{g}}$ with $1\leq i\leq \tilde{%
\ell}=$ rank$\mathbf{\tilde{g}}$. We then argue that scattering matrices of
the general form 
\begin{equation}
S_{\alpha _{(a,i)}\beta _{(b,j)}}^{\gamma _{(b,j)}\delta _{(a,i)}}(\theta
,\sigma _{ij})=\hat{S}_{ab}^{ij}(\theta ,\sigma _{ij})\check{S}_{\alpha
_{a}\beta _{b}}^{\gamma _{b}\delta _{a}}(\theta )  \label{anh}
\end{equation}%
will also satisfy all the above mentioned constraints and constitute
therefore consistent scattering matrices which are by construction invariant
under supersymmetry and allow unstable particles in their spectrum. 

The general formula for $\mathbf{g|\tilde{g}}$-scattering matrices in form
of an integral representation \cite{S5} is 
\begin{eqnarray}
\hat{S}_{ab}^{ij}(\theta ,\sigma _{ij}) &=&\eta _{ab}^{ij}\exp \int_{-\infty
}^{\infty }\frac{dt}{t}\hat{\Phi}(t,h)e^{-it(\theta +\sigma _{ij})},\qquad
\eta _{ab}^{ij}=\exp \left( i\pi \varepsilon _{ij}[K^{-1}]_{\bar{a}b}\right) 
\label{11} \\
\hat{\Phi}_{ab}^{ij}(t) &=&\delta _{ab}\delta _{ij}-\left( 2\cosh \frac{\pi t%
}{h}-\tilde{I}\right) _{ij}\left( 2\cosh \frac{\pi t}{h}-I\right) _{ab}^{-1}.
\label{ijint}
\end{eqnarray}%
Here we denote by $I$ $(\tilde{I})$ and $K$ $(\tilde{K})$ the incidence and
Cartan matrices for the simply laced $\mathbf{g}$ $\mathbf{(\tilde{g})}$-Lie
algebra, respectively. The Coxeter number of $\mathbf{g}$ $\mathbf{(\tilde{g}%
)}$ is $h$ $(\tilde{h})$ and $\varepsilon _{ij}=-\varepsilon _{ji}$ is the
Levi-Civita pseudo-tensor. The special cases $\mathbf{A}_{\ell }\mathbf{|%
\tilde{g}}$ and $\mathbf{g|A}_{1}$ correspond to the $\mathbf{\tilde{g}}%
_{\ell +1}$-homogeneous sine-Gordon models \cite{S16} and $\mathbf{g}$%
-minimal affine Toda field theories (see e.g. \cite{Braden} for a complete
list), respectively. In the ultraviolet limit these models reduce to
conformal field theories, which were discussed in \cite{DHS}, possessing
Virasoro central charges $c^{\mathbf{g|\tilde{g}}}=\ell \tilde{\ell}\,\tilde{%
h}/(h+\tilde{h})$. Besides simply laced Lie algebras, we will here also
allow $\mathbf{g}$ and $\mathbf{\tilde{g}}$ to be the twisted algebra $%
A_{2\ell }^{(2)}$. These cases have not been considered previously. We have
verified here for various examples that the previous formula for the Viasoro
central charge also applies when including $A_{2\ell }^{(2)}$ with rank $%
\ell $ and $h=2\ell +1$ (see below).

The novel feature in $S$-matrices of the type (\ref{11}) is the occurrence
of the resonance parameters $\sigma _{ij}=-\sigma _{ji}$. Besides the first
order poles in the physical sheet which can be interpreted as bound states
of stable particles, there are also simple order poles in the second Riemann
sheet at $\theta _{ab}=-i\eta _{ab}^{\bar{c}}+\sigma _{ab}^{\bar{c}}$ with $%
\eta _{ab}^{\bar{c}}$, $\sigma _{ab}^{\bar{c}}\in \mathbb{R}^{+}$. Poles of
this type admit an interpretation as unstable particles of type $\bar{c}$
with finite lifetime $\tau _{\bar{c}}$. The relations between the masses of
the stable particles $m_{a}$, $m_{b}$, the mass of the unstable particle $m_{%
\bar{c}}$ and the fusing angles $\eta _{ab}^{\bar{c}}$, $\sigma _{ab}^{\bar{c%
}}$ are the Breit-Wigner equations \cite{BW} 
\begin{eqnarray}
m_{\bar{c}}^{2}{}-1/(4\tau _{\bar{c}}^{2})
&=&m_{a}^{2}{}+m_{b}^{2}{}+2m_{a}m_{b}\cosh \sigma _{ab}^{\bar{c}}\cos \eta
_{ab}^{\bar{c}},  \label{BW1} \\
m_{\bar{c}}/\tau _{\bar{c}} &=&2m_{a}m_{b}\sinh \sigma _{ab}^{\bar{c}}\sin
\eta _{ab}^{\bar{c}}\,\,.  \label{BW2}
\end{eqnarray}

The ansatz (\ref{anh}) for the choices $\mathbf{g}=\{A_{2\ell
}^{(2)},A_{\ell }^{(1)},D_{\ell }^{(1)},(C_{\ell }^{(1)}|D_{\ell
+2}^{(2)})\} $ and $\mathbf{\tilde{g}}$ being any simple Lie algebra
satisfies all consistency conditions, in particular the bootstrap equation.
When including non-simply laced Lie algebras, but $A_{2\ell }^{(2)}$, we
also need to take some modifications into account \cite{S15}, from which we
refrain here in order to keep the notation simple.

\section{The ultraviolet limit, a TBA analysis}

Let us now carry out the ultraviolet limit by means of a thermodynamic Bethe
ansatz (TBA) analysis \cite{ZamoTBA} for the above mentioned scattering
matrices by following the work of Ahn \cite{CA}, Moriconi and Schoutens \cite%
{Marco}. In general the TBA is technically very complicated when involving
non-diagonal $S$-matrices. Fortunately, for the case at hand matters
simplify drastically due to the fact that $\hat{S}$ satisfies the so-called
free fermion condition \cite{CA,Marco} (this is a rather misleading
terminology as we are evidently not dealing with free fermions). Here we are
only interested in the extreme ultraviolet limit for which there exists a
standard analysis \cite{ZamoTBA}, which can be adapted to the supersymmetric
case \cite{Marco}. We restrict the following analysis to the ansatz (\ref%
{anh}) for a $\mathbf{g|\tilde{g}}$-supersymmetric $S$-matrix with unstable
particles where we take $\mathbf{g}=A_{2\ell }^{(2)}$. Following \cite%
{CA,Marco} it is straightforward to derive the constant TBA equation for the 
$S$-matrix (\ref{anh}) with the quoted choice of the algebras and all
resonance parameters $\sigma _{ij}$ set to zero 
\begin{equation}
x_{a}^{i}=(1+x_{0}^{i})^{\check{M}_{a}}\prod\limits_{j=1}^{\tilde{\ell}%
}\prod\limits_{b=1}^{\ell }(1+x_{b}^{j})^{N_{ab}^{ij}}\text{,\quad }%
x_{0}^{i}=\prod\limits_{b=1}^{\ell }(1+x_{b}^{i})^{\check{M}_{b}}\text{ \ \
\ for }1\leq i\leq \tilde{\ell},1\leq a\leq \ell .\text{\quad }  \label{TBA}
\end{equation}
The matrices in (\ref{TBA}) are computed from 
\begin{eqnarray}
\hat{N}_{ab}^{ij} &=&\frac{1}{2\pi }\int_{-\infty }^{\infty }d\theta \hat{%
\Phi}_{ab}^{ij}(\theta )=\delta _{ab}\delta _{ij}-\min (a,b)\tilde{K}_{ij},
\label{NN2} \\
\check{N}_{ab} &=&\frac{1}{2\pi }\int_{-\infty }^{\infty }d\theta \check{\Phi%
}_{ab}(\theta )=\frac{1}{2},\quad \quad \quad \text{ }  \label{NN3} \\
\check{M}_{a} &=&\frac{1}{2\pi }\int_{-\infty }^{\infty }d\theta \varphi
_{a}(\theta )=1,\quad \quad  \label{NN4} \\
N_{ab}^{ij} &=&\hat{N}_{ab}^{ij}+\check{N}_{ab}-\frac{1}{2}\check{M}_{a}%
\check{M}_{b}=\delta _{ab}\delta _{ij}-\min (a,b)\tilde{K}_{ij},  \label{NN1}
\end{eqnarray}
with kernels 
\begin{eqnarray}
\check{\Phi}_{ab}(\theta ) &=&\func{Im}\frac{\partial }{\partial \theta }\ln %
\left[ \frac{g_{ab}(\theta )}{\sinh \theta }\right] ,  \label{phi} \\
\varphi _{a}(\theta ) &=&2\func{Im}\frac{\partial }{\partial \theta }\ln %
\left[ \sinh \frac{1}{2}\left( \theta -\frac{i\pi a}{h}\right) \cosh \frac{1%
}{2}\left( \theta +\frac{i\pi a}{h}\right) \right] .  \label{phi2}
\end{eqnarray}
The expression (\ref{NN2}) results directly from (\ref{11}), (\ref{ijint})
when noting that the entries of inverse Cartan matrix of $A_{2\ell }^{(2)}$
are $\min (a,b)$ with $1\leq a,b\leq \ell $. From the solution for the
function $g(\theta )$ in the form (\ref{gint}) we compute the constant (\ref%
{NN3}). Note that the final answer does not depend on the quantities $\eta
_{ij}$, $\hat{\eta}_{ij}$ which were introduced in (\ref{para}) to
parameterize the masses and the constant $\kappa $. The constants $\check{M}%
_{a}$ are obtained by direct computation and hold for all $1\leq a\leq \ell $%
. Assembling then all quantities in (\ref{NN1}) one observes that all
contributions resulting from the supersymmetric factor of the $S$-matrix
have cancelled out, such that $N_{ab}^{ij}=\hat{N}_{ab}^{ij}$. Hence (\ref%
{TBA}) resembles very closely the conventional, that is non-supersymmetric,
constant TBA equations with the modification of the factor involving an
additional particle, named $0$, which results from the diagonalization
procedure.

Having solved (\ref{TBA}) one can compute the effective Virasoro central
charge as 
\begin{equation}
c_{\text{eff}}=\frac{6}{\pi ^{2}}\sum\limits_{k=0}^{\ell }\sum\limits_{j=1}^{%
\tilde{\ell}}\left[ \mathcal{L}\left( \frac{x_{k}^{j}}{1+x_{k}^{j}}\right) -%
\mathcal{L}\left( \frac{y_{k}^{j}}{1+y_{k}^{j}}\right) \right] ,
\label{ceff}
\end{equation}
with $\mathcal{L}\left( x\right) $ denoting Rogers dilogarithm $\mathcal{L}%
(x)=\sum_{n=1}^{\infty }x^{n}/n^{2}+\ln x\ln (1-x)/2$, $x_{a}^{i}=\exp
(-\varepsilon _{a}^{i}(0))$, $y_{a}^{i}=\lim\nolimits_{\theta \rightarrow
\infty }\exp (-\varepsilon _{a}^{i}(\theta ))$ and $\varepsilon
_{a}^{i}(\theta )$ being the rapidity dependent pseudo-energies.

\subsection{The $\mathbf{A}_{2\ell }^{(2)}|\mathbf{\tilde{g}}$-theories}

Even though our main goal is to discuss the supersymmetric scenario, we
shall comment first on the solutions of (\ref{TBA}) and the subsequent
computation of $c_{\text{eff}}$ in the absence of supersymmetry involving
the twisted algebra $\mathbf{A}_{2\ell }^{(2)}$. The reason for this is that
this case will be needed below and has hitherto not been dealt with in the
literature. The supersymmetry is formally broken in (\ref{TBA}) when taking
the limit $\check{M}_{a}\rightarrow 0$ for all $a\in \{1,\ldots ,\ell \}$.
Selecting the algebra which encodes the unstable particles to be $\mathbf{%
\tilde{g}}=A_{\tilde{\ell}}^{(1)}$, we found the following analytic
solutions for (\ref{TBA}) 
\begin{equation}
x_{a}^{j}=\frac{\sin j\pi /\tau \sin (\tilde{h}-j)\pi /\tau }{\sin a\pi
/\tau \sin (\tilde{h}+a)\pi /\tau }\qquad \text{for \ }1\leq a\leq \ell 
\text{, }1\leq j\leq \tilde{\ell},  \label{xA}
\end{equation}%
where $h$ $(\tilde{h})$ is the Coxeter number of $A_{2\ell }^{(2)}$ $(A_{%
\tilde{\ell}}^{(1)})$, namely $h=2\ell +1$ $(h=2\tilde{\ell}+1)$ and $\tau
=h+\tilde{h}$. In fact, the solutions (\ref{xA}) for the constant
TBA-equations (\ref{TBA}) hold for all four $\mathbf{g|\tilde{g}}$-theories
with $\mathbf{g,\tilde{g}\in \{}A_{n}^{(1)},A_{2n}^{(2)}\}$. Computing the
effective central charge by means of (\ref{ceff}) yields the usual value 
\cite{S5} of the $\mathbf{g|\tilde{g}}$-theories 
\begin{equation}
c_{\text{eff}}=\frac{\ell \tilde{\ell}\tilde{h}}{h+\tilde{h}}.  \label{ceff2}
\end{equation}%
We have solved (\ref{TBA}) for other $\mathbf{A}_{2\ell }^{(2)}|\mathbf{%
\tilde{g}}$-theories involving various simply laced algebras $\mathbf{\tilde{%
g}}$ and obtained (\ref{ceff2}) in all cases. So far we have not found
simple closed expressions for the $x_{a}^{j}$ as in (\ref{xA}) for these
cases and will not present here more case-by-case results.

\subsection{The $N=1$ supersymmetric $\mathbf{A}_{2\ell }^{(2)}|\mathbf{%
\tilde{g}}$-theories}

We shall now turn to the full supersymmetric version of the constant
TBA-equations (\ref{TBA}) describing the $\mathbf{A}_{2\ell }^{(2)}|\mathbf{%
\tilde{g}}$-theories. In general, we may write (\ref{TBA}) as 
\begin{equation}
x_{a}^{i}=(1+x_{0}^{i})\prod\limits_{j=1}^{\tilde{\ell}%
}(1+x_{1}^{j})^{N_{a1}^{ij}}\prod\limits_{j=1}^{\tilde{\ell}%
}\prod\limits_{b=2}^{\ell }(1+x_{b}^{j})^{N_{ab}^{ij}}\text{\ \ ~\quad\ for }%
1\leq i\leq \tilde{\ell},1\leq a\leq \ell .  \label{1}
\end{equation}%
Excluding $a=1$ and noting the simple fact that $\min (a,b)=\min (a-1,b-1)+1$
we may re-write (\ref{1}) for the theories at hand as 
\begin{equation}
x_{a}^{i}=(1+x_{0}^{i})\prod\limits_{j=1}^{\tilde{\ell}%
}(1+x_{1}^{j})^{-K_{ij}}\prod\limits_{b=2}^{\ell
}(1+x_{b}^{j})^{-K_{ij}}(1+x_{b}^{j})^{N_{(a-1)(b-1)}^{ij}}\text{\ \ ~ for }%
2\leq a\leq \ell .
\end{equation}%
Taking the limit $x_{1}^{i}\rightarrow \infty $ \ of this equation leads to 
\begin{equation}
x_{a}^{i}=\lim_{x_{1}^{i}\rightarrow \infty }\left[ x_{1}^{i}\prod%
\limits_{j=1}^{\tilde{\ell}}(x_{1}^{j})^{-K_{ij}}\prod\limits_{b=2}^{\ell
}(1+x_{b}^{i})(1+x_{b}^{j})^{-K_{ij}}(1+x_{b}^{j})^{N_{(a-1)(b-1)}^{ij}}%
\right] ~.  \label{lim}
\end{equation}%
For $\mathbf{\tilde{g}}=A_{2}^{(1)}$ we may assume that $x_{a}^{1}=x_{a}^{2}$%
, such that (\ref{lim}) simplifies to 
\begin{equation}
x_{a}^{i}=\prod\limits_{j=1}^{\tilde{\ell}}\prod\limits_{b=2}^{\ell
}(1+x_{b}^{j})^{N_{(a-1)(b-1)}^{ij}}~\quad \text{\ for }1\leq i\leq \tilde{%
\ell},2\leq a\leq \ell .
\end{equation}%
which is precisely the system for an $\mathbf{A}_{2(\ell
-1)}^{(2)}|A_{2}^{(1)}$-theory when renaming the particles $a=2$ to $a=1$, $%
a=3$ to $a=2$, $\ldots $, $a=\ell $ to $a=\ell -1$. The solutions for the
constant TBA equations (\ref{TBA}) are therefore in this case $x_{0}^{1}=$ $%
x_{0}^{2}=x_{1}^{1}=x_{2}^{1}\rightarrow \infty $ and $x_{k}^{1}=$ $x_{k}^{2}
$ for $1\leq k\leq \ell -1$ given by (\ref{xA}). Taking further $%
y_{0}^{1}=y_{0}^{2}=1$, \ $y_{k}^{1}=\ y_{k}^{2}=0$ for $1\leq k\leq \ell -1$
the effective central charge is then computed by means of (\ref{ceff}) 
\begin{equation}
c_{\text{eff}}=\frac{6}{\pi ^{2}}\left[ \sum\limits_{k=1}^{\ell
-1}\sum\limits_{j=1}^{2}\mathcal{L}\left( \frac{x_{k}^{j}}{1+x_{k}^{j}}%
\right) +4\mathcal{L}\left( 1\right) -2\mathcal{L}\left( \frac{1}{2}\right) %
\right] =\frac{3(\ell -1)}{\ell +1}+3=\frac{6\ell }{\ell +1}.
\end{equation}%
We may also investigate the behaviour of these theories for large resonance
parameters. As the bosonic part of the theory decouples in this case into
two separate non-interacting theories \cite{HSGTBA,F5,S1}, the two $N=1$
supersymmetric theories will behave analogously due to the factorization
ansatz for $S$. Accordingly we have 
\begin{equation}
\lim_{\sigma _{12}\rightarrow \infty }\mathbf{A}_{2\ell
}^{(2)}|A_{2}^{(1)}\rightarrow \mathbf{A}_{2\ell }^{(2)}|A_{1}^{(1)}\otimes 
\mathbf{A}_{2\ell }^{(2)}|A_{1}^{(1)}\text{ .}
\end{equation}%
The resulting $\mathbf{A}_{2\ell }^{(2)}|A_{1}^{(1)}$-theories are the $N=1$
supersymmetric theories discussed in \cite{Schoutens,Marco}. The effective
Virasoro central charge is then simply obtained as the sum of the known
effective central charges of the supersymmetric minimal $\mathcal{SM}%
(2,4\ell +4)$ conformal field theories 
\begin{equation}
c_{\text{eff}}=\frac{3\ell }{2\ell +2}+\frac{3\ell }{2\ell +2}=\frac{3\ell }{%
\ell +1}.
\end{equation}%
Thus we observe that the effective central charge for the theory with
vanishing resonance parameter is twice the one with large resonance
parameter.

Clearly it is interesting to carry out the TBA analysis for other algebras $%
\mathbf{\tilde{g}}$. Here it suffices to have demonstrated that the proposed
scattering matrices of the type (\ref{anh}) have a meaningful ultraviolet
behaviour, which in the presented cases can even be obtained analytically.

\section{Conclusion}

We have shown that our $S$-matrix proposal (\ref{anh}) consistently combines 
$N=1$ supersymmetry with the requirement to have unstable particle in the
spectrum of the theory. The $S$-matrix satisfies all the constraints imposed
by the bootstrap program and possesses a sensible ultraviolet limit.

There are various open issues which would be interesting to address in
future. The proposal (\ref{anh}) constitutes the first concrete example for
a non-diagonal scattering matrix corresponding to a theory which contains
unstable particles. It would be interesting to construct further
non-diagonal scattering matrices of this type for which the supersymmetry is
broken or possibly enlarged to greater values of $N$.

Clearly it would be interesting to complete the detailed analysis involving
also other algebras on the bosonic side. More challenging is to modify the
boson-fermion mixing part. So far the entire fusing structure of the model
was confined to the bosonic factor. However, we also provided solutions for
the function $g(\theta )$ which has simple poles in the physical sheet,
which can be interpreted as stable bound states possibly leading to
consistent solutions for the bootstrap equations. Concerning the
implementation of unstable particles, it should also be possible to extend
the ansatz (\ref{anh}) to the form 
\begin{equation}
S_{\alpha _{(a,i)}\beta _{(b,j)}}^{\gamma _{(b,j)}\delta _{(a,i)}}(\theta
,\sigma _{ij})=\hat{S}_{ab}^{ij}(\theta ,\sigma _{ij})\check{S}_{\alpha
_{(a,i)}\beta _{(b,j)}}^{\gamma _{(b,j)}\delta _{(a,i)}}(\theta ,\sigma
_{ij})~.
\end{equation}%
This would means that the unstable particles are be no longer of a purely
bosonic nature. More general alterations such as taking the bosonic factor
to be non-diagonal or relaxing the factorzation ansatz into a purely bosonic
and boson-fermion mixing factor altogether have not even been considered in
the absence of unstable particles.

\bigskip \noindent \textbf{Acknowledgments:} I am grateful to M. Moriconi
for useful discussions and to the members of the Institut f\"{u}r
theoretische Physik at the University of Hannover for kind hospitality.


\end{document}